\documentclass[letter]{aa} 

\usepackage{txfonts}
\usepackage{hyperref}

\usepackage{graphicx}   
\usepackage{amsmath}    
\usepackage{amssymb}    

\begin{document} 

\title{Discovery of the first heavily obscured QSO candidate at $z>6$ in a close galaxy pair}
\titlerunning{The first heavily obscured QSO at $z>6$}
\authorrunning{F. Vito et al.}
   \author{F. Vito\thanks{fabio.vito@uc.cl}\inst{1,2} \and
          W.N. Brandt\inst{3,4,5} \and
          F.E. Bauer\inst{1,6,7} \and
          R. Gilli\inst{8} \and
          B. Luo\inst{9,10,11} \and
          G. Zamorani\inst{8} \and
          F. Calura\inst{8} \and
          A. Comastri\inst{8} \and
          C. Mazzucchelli\inst{12} \and
          M. Mignoli\inst{8} \and
          R. Nanni\inst{8,13} \and
          O. Shemmer\inst{14} \and
          C. Vignali\inst{8,13} \and
          M. Brusa\inst{8,13} \and
          N. Cappelluti\inst{15} \and
          F. Civano\inst{16} \and
          M. Volonteri\inst{17}
          }
 \institute{Instituto de Astrof\'isica and Centro de Astroingenieria, Facultad de F\'isica, Pontificia Universidad Cat\'olica de Chile, Casilla 306, Santiago 22, Chile
         \and
Chinese Academy of Sciences South America Center for Astronomy, National Astronomical Observatories, CAS, Beijing 100012, China
\and
Department of Astronomy \& Astrophysics, 525 Davey Lab, The Pennsylvania State University, University Park, PA 16802, USA
\and
Institute for Gravitation and the Cosmos, The Pennsylvania State University, University Park, PA 16802, USA
\and
Department of Physics, The Pennsylvania State University, University Park, PA 16802, USA
\and
Millennium Institute of Astrophysics (MAS), Nuncio Monse\~nor S\'otero Sanz 100, Providencia, Santiago, Chile
\and
Space Science Institute, 4750 Walnut Street, Suite 205, Boulder, Colorado, 80301, USA
\and
INAF -- Osservatorio di Astrofisica e Scienza dello Spazio di Bologna, Via Gobetti 93/3, I-40129 Bologna, Italy
\and
School of Astronomy and Space Science, Nanjing University, Nanjing 210093, China
\and
Key Laboratory of Modern Astronomy and Astrophysics, Nanjing University, Ministry of Education, Nanjing, Jiangsu 210093, China
\and
Collaborative Innovation Center of Modern Astronomy and Space Exploration, Nanjing 210093, China
\and
European Southern Observatory, Alonso de C\'ordova 3107, Vitacura, Regi\'on Metropolitana, Chile
\and
Dipartimento di Fisica e Astronomia, Universit\`a degli Studi di Bologna, via Gobetti 93/2, I-40129 Bologna, Italy
\and
Department of Physics, University of North Texas, Denton, TX 76203, USA
\and
Physics Department, University of Miami, Coral Gables, FL 33124, USA
\and
15 Center for Astrophysics | Harvard \& Smithsonian, 60 Garden st, Cambridge, MA 02138, USA
\and
Sorbonne Universit\'es, UPMC Universit\'e Paris 06 et CNRS, UMR7095, Institut d'Astrophysique de Paris, 98bis boulevard Arago, \\F-75014, Paris, France
  }

   \date{}
  \abstract
{While theoretical arguments predict that most of the early growth of supermassive black holes (SMBHs) happened during heavily obscured phases of accretion, current methods used for selecting $z>6$ quasars (QSOs) are strongly biased against obscured QSOs, thus considerably limiting our understanding of accreting SMBHs during the first gigayear of the Universe from an observational point of view. We report the \textit{Chandra} discovery of the first heavily obscured QSO candidate in the early universe, hosted by a close ($\approx5$ kpc) galaxy pair at $z=6.515$. One of the members is an optically classified type-1 QSO, PSO167--13. The companion galaxy was first detected as a \mbox{[C II]} emitter by Atacama large millimeter array (ALMA). An X-ray source is significantly ($P=0.9996$) detected by \textit{Chandra} in the 2--5 keV band, with $<1.14$ net counts in the 0.5--2 keV band, although the current positional uncertainty does not allow a conclusive association with either PSO167--13 or its companion galaxy. From \mbox{X-ray} photometry and hardness-ratio arguments, we estimated an obscuring column density of $N_H>2\times10^{24}\,\mathrm{cm^{-2}}$ and $N_H>6\times10^{23}\,\mathrm{cm^{-2}}$ at $68\%$ and $90\%$ confidence levels, respectively. Thus, regardless of which of the two galaxies is associated with the X-ray emission, this source is the first heavily obscured QSO candidate at $z>6$.}


   \maketitle
%

\section{Introduction}

\begin{figure}
        \begin{center}
                \hbox{
                        \includegraphics[width=90mm,keepaspectratio]{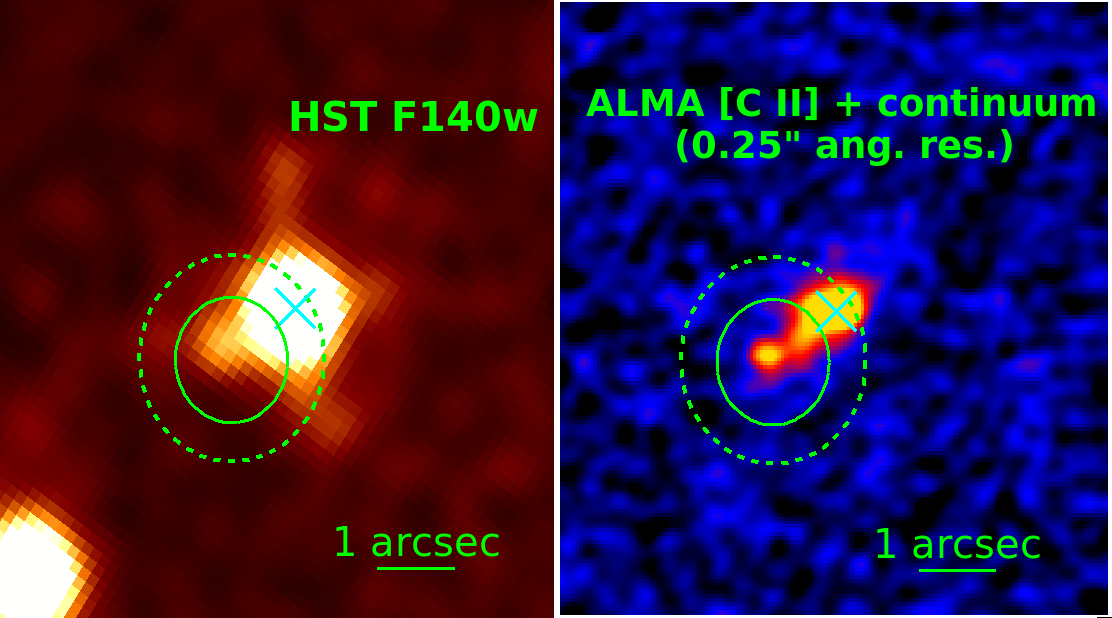} 
                }
        \end{center}
        \caption{ 
                \textit{HST} and high-resolution ($\approx0.25''$) ALMA images \citep{Mazzucchelli19, Neeleman19}, revealing the companion galaxy in the F140W band and [C II] line, respectively. The green solid and dashed circles show the X-ray positional uncertainty corresponding to 68\% and 90\% confidence levels (0.73 arcsec and 1.17 arcsec, respectively), and are centered on the hard-band emission centroid (see Fig.~\ref{Xray}).  The cyan cross marks the PanSTARRS-1 position of PSO167--13.}\label{J1110}
\end{figure}

The discovery of accreting supermassive black holes (SMBHs) with masses of $10^9-10^{10}\,M_\odot$ shining as quasars (QSOs) at $z>6$ \citep[e.g.,][]{Mortlock11, Banados16,Banados18a} when the Universe was less than 1 Gyr-old challenges our understanding of SMBH formation and growth in the early universe, and is one of the major open issues in modern astrophysics \citep[e.g.,][]{Reines16,Woods18}. 
Different classes of theories have been proposed to explain the formation of the BH seeds that eventually became SMBHs. The two most popular classes of models involve the formation of ``light seeds'' ($M \approx 10^2\,M_\odot$), as remnants of the first Pop III stars, and ``heavy seeds" ($M \approx 10^4-10^6\,M_\odot$), perhaps formed during the direct collapse of giant pristine gas clouds (e.g., \citealt{Volonteri16b, Valiante17, Smith18,Woods18}, and references therein). 
To match the masses of the SMBHs discovered at $z>6$, all such models require continuous, nearly Eddington-limited or even super-Eddington accretion phases during which the growing SMBH is expected to be heavily obscured by the same accreting material with large column densities, even exceeding the Compton-thick level ($N_H=1.5\times10^{24}\,\mathrm{cm^{-2}}$; e.g., \citealt{Pacucci15,Pezzulli17}). ``Wet'' (i.e., gas-rich) galaxy mergers are expected to provide both a large amount of gas and the mechanisms to drive it toward the galaxy nuclear regions, thus allowing efficient SMBH accretion \citep[e.g.,][but see also \citealt{DiMatteo12}]{Hopkins08}. 
Indeed, high-redshift QSOs are usually found in overdense environments in simulations \citep[e.g.,][]{Costa14,Barai18,Habouzit18}, but no consensus has yet been reached among observational works \citep[e.g.,][]{Balmaverde17,Mazzucchelli17a, Ota18}.

Currently, approximately 180 quasars have been discovered at $z> 6$ (e.g., \citealt{Banados16} and references therein; \citealt{Mazzucchelli17b, Matsuoka18a,Matsuoka19,Wang18a, Fan19,Reed19}), up to $z = 7.54$ (ULAS J1342+0928; \citealt{Banados18a}). 
However, these rare QSOs have been selected from wide-field optical/near-infrared(NIR) surveys such as, for example, SDSS, CFHQS, and PanSTARRS-1, and thus are, by selection, optically type~1 (i.e., broad emission-line QSOs with blue UV continua). The selection of $z>6$ QSO candidates typically relies on the detection of the blue power-law UV continuum, absorbed at $\lambda<1216$ \AA\, by the $Ly\alpha$ forest, and suppressed at wavelengths shorter than the $Ly\alpha$ break at 912 \AA, due to absorption by intervening neutral hydrogen. Therefore, the census of accreting SMBHs in the early universe is currently missing, by selection, the key population of obscured systems, 
thereby strongly limiting our understanding of the early phases of SMBH growth. Currently, the highest redshift, Compton-thick QSO candidate is XID403 at $z=4.76$ \citep{Gilli14,Circosta19}, an X-ray-selected QSO in the \textit{Chandra} Deep Field-South \citep{Xue11,Luo17}. 

In this Letter, we report the discovery in the X-ray band of the first heavily obscured QSO candidate at $z>6$, in a close ($\approx5$ kpc) pair of galaxies at $z\approx6.515$. One of the two galaxies hosts an optically classified type-1 QSO, PSO J167.6415--13.4960 (hereafter PSO167--13). Evidence for interaction between the two galaxies is reported in \cite{Mazzucchelli19}.
Errors and limits are reported at the $68\%$ confidence level, unless otherwise noted. We adopt a flat cosmology with $H_0=67.7\,\mathrm{km\,s^{-1}}$ and $\Omega_m=0.307$ \citep{Planck16}.

\begin{figure*}
                \hbox{
                        \includegraphics[width=180mm,keepaspectratio]{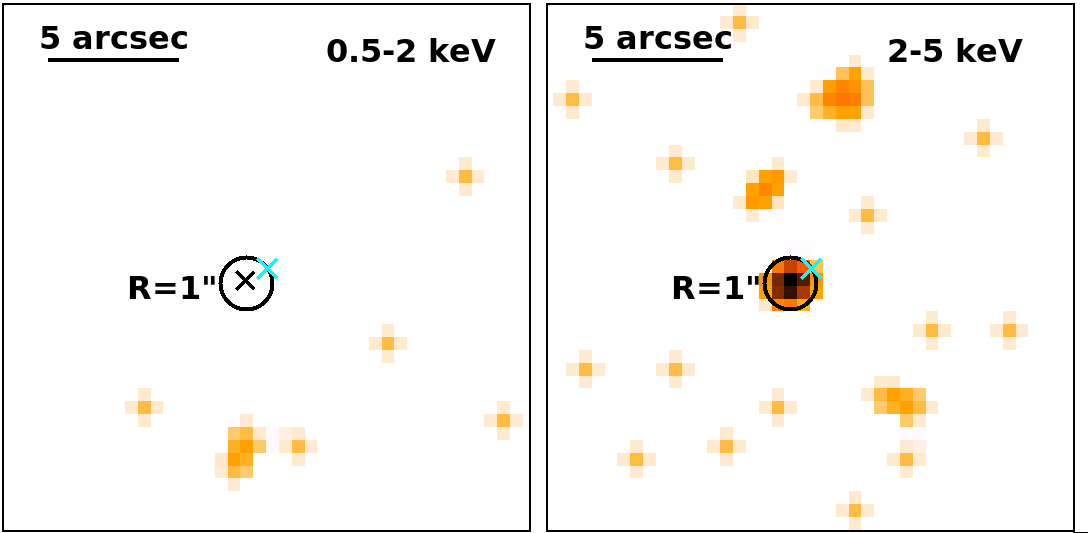} 
                }
        \caption{Soft (\textit{left panel}) and hard (\textit{right panel}) band \textit{Chandra} images of a $20''\times20''$ region around the UV position of PSO167--13 (cyan cross), smoothed with a Gaussian function with a three-pixel kernel radius. The black cross (not shown in the right panel for clarity) marks the \textit{ALMA} position of the companion galaxy (Fig.~\ref{J1110}, right panel).
                The X-ray source is significantly ($P=0.9996$) detected in the hard band, while no counts are detected in the soft band. The black circle ($R=1$ arcsec) is centered on the hard-band emission centroid and is used for photometry evaluation. }\label{Xray}
\end{figure*}

\section{Target description and data analysis}
PSO167-13 was first selected as a high-redshift QSO candidate on the basis of its colors in the PanSTARRS-1 survey (\citealt{Venemans15a}, see Fig.~\ref{J1110}, left panel), and was then confirmed spectroscopically to lie at $z=6.515$ both in the rest-frame UV \citep{Venemans15a} and sub-millimeter with Atacama large millimeter array (ALMA), via detection of the [C II] ($158\,\mu m$) emission line \citep{Decarli18}. An investigation of the ALMA data-cube at frequencies near the [C II] emission line  \citep{Willott17} revealed the presence of a close companion, separated by $0.9''$ ($\approx5$ kpc in projection at the redshift of the QSO) from the rest-frame UV and [C II] position of the QSO, and by $\Delta v\approx-270\,\mathrm{km\,s^{-1}}$ (corresponding to $\Delta z \approx 0.007$) in velocity space (based on the frequency of the [C II] emission peaks). The companion galaxy thus forms a physical pair with PSO167--13. Its existence was recently confirmed by a deep HST/WFC3 observation in the F140W 
($\approx1.4\,\mu m$) band (with AB magnitude F140W$=25.5$) and new high-resolution ($\approx0.25"$) ALMA imaging (Fig.~\ref{J1110}, center and right panels; \citealt{Mazzucchelli19, Neeleman19}). No rest-frame UV spectrum is currently available for this galaxy. 
Similar companions have been found in about a quarter of the $z>6$ QSOs observed with ALMA \citep{Decarli17}.

We observed PSO167--13 for 59 ks with \textit{Chandra} as part of a larger program aimed at making exploratory observations of a statistically significant sample of ten $z>6$ QSOs (Vito et al., in prep.).\footnote{\textit{ Chandra} observations of the remaining 9 targets have been completed and the analysis is ongoing. PSO167--13 is the only source showing significant evidence of obscuration.}
We reprocessed the \textit{Chandra} observations with the \textit{chandra\_repro} script in CIAO 4.10,\footnote{http://cxc.harvard.edu/ciao/} using CALDB v4.8.1,\footnote{http://cxc.harvard.edu/caldb/} setting the option \textit{check\_vf\_pha=yes} in the case of observations taken in Very Faint mode, and extracted the response matrix and ancillary file using the \textit{specextract } tool. 
The astrometry for all instruments has been consistently locked on the PanSTARRS-1 frame, using six common sources in the field for \textit{Chandra} (we used the CIAO \textit{wcs\_match} and \textit{wcs\_update} tools), and the position of PSO167--13 itself for HST and ALMA.

We detected significant emission in the hard (2--5 keV) band using a standard circular extraction region of $1$ arcsec radius (Fig.~\ref{Xray}). In particular, we detected three counts, with an expected background level of 0.14 counts, corresponding to a number of net counts of $2.86_{-1.44}^{+2.14}$ and a false-detection probability (i.e., that the detected emission is due to a background fluctuation) of only $P=4\times10^{-4}$ \citep{Weisskopf07}. The corresponding flux is $F_{2-5\,\mathrm{keV}}=8.1_{-3.9}^{+5.9}\times10^{-16}\,\mathrm{erg\,cm^{-2}\,s^{-1}}$. 
As a check on the detection significance, after having masked bright sources  including PS167--13, we performed aperture photometry using $R=1$ arcsec regions randomly centered over $10^5$ positions across the field in the $2-5$ keV band, and detected $\geq3$ counts for 52 of them ($P=5\times10^{-4}$). Moreover, 10 of these 52 regions are also coincident with the positions of PanSTARRS galaxies, and therefore could be real X-ray sources, increasing the agreement with the false-source probability reported above. All of the three detected counts have energies in the range $2.5\lesssim \frac{E}{\mathrm{keV}}\lesssim 3.5$, which is not surprising since the effective area of  \textit{Chandra} drops at high energies. If we restrict the detection to the $2-4$ keV band, thus excluding the background-dominated higher energies, we derive an even higher detection significance ($P=2\times10^{-4}$). 

We detected zero counts in the soft (0.5--2 keV) band at the UV position of PSO167--13 (cyan cross in Fig.~\ref{Xray}), thus setting an upper limit on the net counts of $<1.14$ \mbox{\citep{Weisskopf07}}. In order to evaluate the significance of the soft-band nondetection, we assumed a standard $\Gamma=1.9$ power-law model, suitable for high-redshift luminous QSOs (e.g., 
\mbox{\citealt{Shemmer06b}}, \citealt{Nanni17}), normalized to the observed net-count rate in the hard band. Accounting for Galactic absorption, the expected background in the extraction region ($\approx0.1$ counts), and the \textit{Chandra} effective area at the position of the target, the expected number of soft-band counts is 6.59. Given this expectation, the Poisson probability of detecting zero counts is $P(x=0,\mu=6.59)=1.37\times10^{-3}$. Conservatively assuming a rather flat slope ($\Gamma=1.6$, based on the uncertainties on the average photon index in \citealt{Shemmer06b,Nanni17}), the source nondetection in the soft band remains significant ($P=6.1\times10^{-3}$).

The centroid of the hard-band emission is shifted from the optical and sub-millimeter position of PSO167--13 (cyan cross in Fig.~\ref{Xray}) by $0.97$ arcsec and by $0.15$ arcsec from the [C II] position of the companion galaxy (black cross). We computed the positional uncertainty via 1000 MARX 5.3.3\footnote{https://space.mit.edu/ASC/MARX/} simulations of a source with three counts in the hard band at the position of PSO167--13, accounting for the real instrumental configuration, and including a (negligible) residual astrometry uncertainty. We found a positional uncertainty of $0.73$ arcsec and $1.17$ arcsec at 68\% and 90\% confidence levels, respectively. The observed offset between the X-ray source and the optical position of PSO167--13 is significant at $\approx1.5\sigma$ only, such that the hard-band X-ray emission is consistent with being produced by the type-1 QSO.

\section{Results and discussion}

The measured X-ray photometry corresponds to a hardness ratio\footnote{$HR=(H-S)/(H+S)$, where $S$ and $H$ are source net counts in the soft and hard bands, respectively. This quantity is widely used to characterize the spectra of X-ray sources with limited photon statistics.} 
of $HR>0.47$ and an effective power-law photon index at rest-frame $4-38$ keV of $\Gamma<0.55$, computed accounting for the effective area at the position of the X-ray source and Galactic absorption. These extremely hard values for an object at $z=6.515$ strongly suggest the source is heavily obscured. We estimated the column density required to retrieve such values through spectral simulations with XSPEC, assuming an intrinsic power-law spectrum with $\Gamma=1.9$ and accounting for Galactic absorption, and obtained $N_H>2\times10^{24}\,\mathrm{cm^{-2}}$ and $N_H>6\times10^{23}\,\mathrm{cm^{-2}}$ at the $68\%$ and $90\%$ confidence levels, respectively. The column density cannot be constrained at $\gtrsim95\%$ confidence level, due to the combination of the number of detected counts and the photoelectric cut-off shifting outside the \textit{Chandra} band for low column densities. 

We estimated the rest-frame $2-10$ keV luminosity of this source from the detected counts in the observed-frame hard band  assuming $\Gamma=1.9$ to be in the range $L_{2-10\mathrm{keV}}=[6.6_{-3.2}^{+4.5}, 7.4_{-3.6}^{+5.4}]\times10^{44}\,\mathrm{erg\,s^{-1}}$, where the lower and upper limits are computed assuming \mbox{$N_H=[0, 2]\times10^{24}\,\mathrm{cm^{-2}}$}, respectively. 
The derived luminosity does not vary significantly for very different values of $N_H$, as the high rest-frame energies (i.e., $15-38$ keV) probed at $z=6.52$ are not strongly affected by even moderately Compton-thick obscuration.  

\begin{figure}
        \begin{center}
                \includegraphics[width=80mm,keepaspectratio]{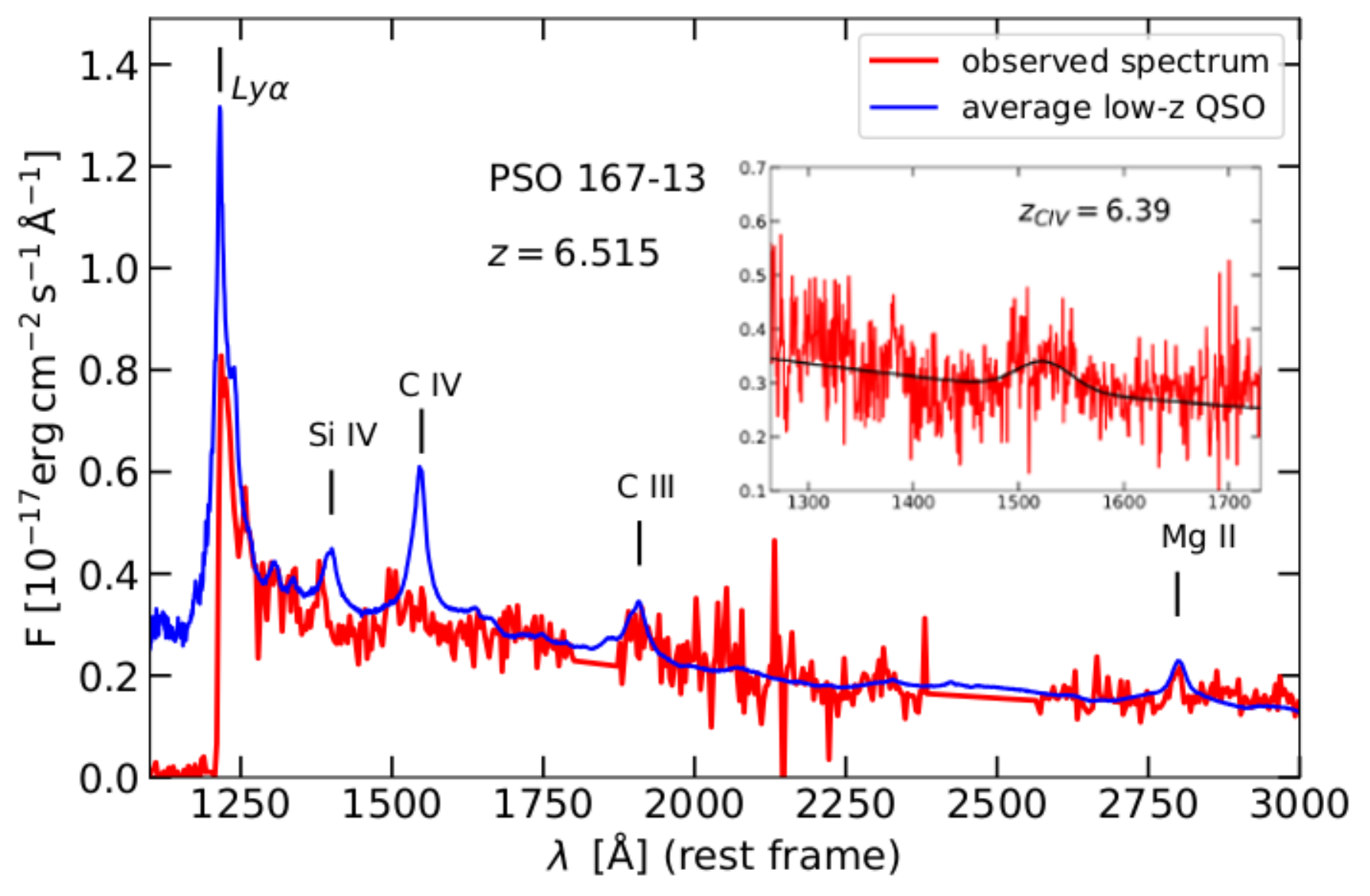} 
        \end{center}
        \caption{Rest-frame UV spectrum of PSO167-13 (red line, adapted from \citealt{Mazzucchelli17b}) obtained with VLT/FORS2 and Magellan/FIRE, compared with the average QSO spectrum of \citet[blue line]{VandenBerk01}. A zoom in the C IV region is shown in the inset, where we also report the best fit to the C IV line, and associated redshift and equivalent width. Tentative BAL features might be present at wavelengths bluer than the \mbox{C IV} emission line, which is also particularly weak. PSO167--13 may thus be a BAL QSO and/or a WLQ.}\label{spectrum}
\end{figure}

Considering the hard-band positional accuracy (see the green circle in the 
center and right panels of  Fig.~\ref{J1110},), the source of the X-ray emission could be either PSO167--13 or its companion galaxy. 
 Assuming that the QSO is the source of the hard-band emission with a somewhat large X-ray offset (0.97 arcsec), the upper limit we derived above on the number of soft-band counts corresponds to an observed
soft-band X-ray emission $\gtrsim4$ times weaker than that expected from its UV luminosity
\citep[e.g.,][]{Just07}. The intrinsic (i.e., corrected for absorption) luminosity  estimated in the previous paragraph would be consistent within a factor of two with the expectations based on the QSO UV luminosity \citep[e.g.,][]{Just07}. Several physical processes could explain why an optically classified type-1 QSO is heavily obscured in the X-rays. For instance, approximately $50\%$ of the Weak Emission-Line QSOs (WLQs; e.g., \citealt{Diamond-Stanic09}) are associated with weak and hard X-ray emission \citep[e.g.,][]{Luo15, Ni18}, possibly linked to the presence of thick accretion disks with large column density on small scales that prevent ionizing radiation from reaching the broad-line region. Moreover, WLQs are usually found to be fast-accreting QSOs \citep[e.g.,][]{Luo15,Marlar18}, as is PSO167--13 ($\lambda_{\mathrm{Edd}}\gtrsim1$; \citealt{Mazzucchelli17b}). 

Similarly, in broad-absorption-line QSOs (BALQSOs), small-scale screening material may absorb the ionizing UV/X-ray radiation, thus allowing the acceleration of the outflowing wind producing the BALs \citep[e.g.,][]{Proga04}, but still allowing the detection of the blue UV continuum. \citet[see also \citealt{Yi19}]{Rogerson18} reported the emergence of BALs on timescales of $\sim100$ days, that is, shorter than the rest-frame time that passed from the UV spectral observation to the X-ray imaging of PSO167--13, which was $6$ months.

In the currently available  rest-frame UV spectrum of PSO167--13 (Fig.~\ref{spectrum}) the C IV line is relatively weak ($EW_{CIV}\approx12\,\mathrm{\AA}$), and some BAL features might also be present at bluer wavelengths. The C IV line is blueshifted by $\sim-5800\,\mathrm{km\,s^{-1}}$ with respect to the Mg II emission line, similarly to some hyper-luminous QSOs \citep[e.g.,][]{Vietri18} and other $z>6$ QSOs \citep[e.g.,][]{Banados18a,Meyer19}. Such extreme blueshifts are also seen in WLQs \citep[e.g.,][]{Luo15,Ni18}.

We also note that \cite{Nanni18} detected significant spectral variability (from $N_H\approx0\,\mathrm{cm^{-2}}$ to \mbox{$N_H\approx5\times10^{23}\,\mathrm{cm^{-2}}$)} in two distinct X-ray observations of the $z>6$ QSO SDSS J1030+0524. A similar increase of the obscuration might have taken place also for \mbox{PSO167--13} between its observations in rest-frame UV and X-rays.  Additional rest-frame UV spectroscopic observations are also needed to investigate such a possibility, as well as to help characterize this object as a possible WLQ or BALQSO. 

Alternatively, since the X-ray centroid is consistent with the position of the companion galaxy, this could host a heavily obscured QSO, in a close and interacting \citep{Mazzucchelli19} pair with PSO167--13.
In this scenario, only its proximity to the optically type-1 QSO allowed us to discover it with \textit{Chandra}, as high-redshift obscured  QSOs are missed by UV surveys, and the lack of strong detection of X-ray emission from PSO167--13 can be explained by a moderate intrinsic X-ray weakness (a factor of $\geq4$). Similar pairs of QSOs have been discovered at redshifts as high as $z\approx5$ \citep{McGreer16}, although with larger separation, but beyond $z\approx3.3$ none are known to include an obscured QSO \citep{Vignali18}. 

 To summarize, if PSO167--13, optically classified as a type-1 QSO, were found to be responsible for the high-energy emission, it would be an intrinsically X-ray normal but heavily obscured QSO, and the causes of the UV/X-ray misclassification would need to be investigated. Alternatively, if the companion galaxy were found to be the X-ray source, it would be a heavily obscured QSO in an interacting pair with PSO167--13. In this case, PSO167--13 would be intrinsically X-ray weak by a factor of $\geq4$. Thus, regardless of which of the two members of the system produced the hard-band detection, it represents the first heavily obscured QSO candidate in the early universe.  
 Deeper X-ray observations are required to better constrain the column density and to improve the positional accuracy of the hard-band X-ray source, thereby allowing a confident association with either \mbox{PSO167--13} or its companion galaxy, and confirmation or rejection of the QSO-pair nature of this system.

\begin{acknowledgements}
We thank the anonymous referee for their useful comments and suggestions.
FV acknowledges financial support from CONICYT and CASSACA through the Fourth call for tenders of the CAS-CONICYT Fund. WNB acknowledges \textit{Chandra} X-ray Center grant G08-19076X. BL acknowledges financial support from the National
Key R\&D Program of China grant 2016YFA0400702 and
National Natural Science Foundation of China grant
11673010. We acknowledge financial contribution from CONICYT grants Basal-CATA AFB-170002 (FV, FEB), 
the Ministry of Economy, Development, and Tourism's Millennium Science
Initiative through grant IC120009, awarded to The Millennium Institute
of Astrophysics, MAS (FEB), and the agreement ASI-INAF n.2017-14-H.O.
\end{acknowledgements}

%
%
\bibliographystyle{aa}
\bibliography{../../../../../biblio.bib} 

%
\end{document}